\documentclass[12pt,preprint]{aastex}

\newcommand{\HI}{\ion{H}{1}\ }              
\newcommand{\HII}{\ion{H}{2}\ }             
\newcommand{\Ha}{\rm H$\alpha$\ }
\newcommand{\h}{^{\rm h}}
\newcommand{\m}{^{\rm m}}
\newcommand{\s}{^{\rm s}}
\newcommand{\kms}{km\,s$^{-1}$}
\newcommand{\Msun}{{M$_{\odot}$}} 
\shorttitle{Triggering and Feedback Processes in NGC~1569}
\shortauthors{M\"uhle et al.}

\begin{document}



\title{Triggering and Feedback: The Relation between the HI Gas and the Starburst in the Dwarf Galaxy NGC~1569}

\author{S. M\"uhle\altaffilmark{1} and U. Klein}
\affil{Radioastronomisches Institut der Univ.\ Bonn, Auf dem H\"ugel 71, 
 53121 Bonn, Germany}
\altaffiltext{1}{currently at the Department of Astronomy and Astrophysics, 
University of Toronto, 60 St.\ George St., Toronto, Ontario, M5S 3H8, Canada, 
muehle@astro.utoronto.ca}
\email{smuehle@astro.uni-bonn.de, uklein@astro.uni-bonn.de}

\author{E.M. Wilcots}
\affil{Washburn Observatory, University of Wisconsin, 475 N. Charter St., 
 Madison, Wisconsin 53706, USA}
\email{ewilcots@uwast.astro.wisc.edu}

\and 

\author{S. H\"uttemeister}
\affil{Astronomisches Institut der Ruhr-Universit\"at Bochum, 
 Universit\"atsstr. 150, 44780 Bochum, Germany}
\email{huette@astro.ruhr-uni-bochum.de}

\begin{abstract}
As part of our study on the impact of violent star formation on the
interstellar medium (ISM) of dwarf galaxies, we report observations of
neutral atomic hydrogen (\ion{H}{1}) in the post-starburst dwarf galaxy 
NGC~1569. 
High-resolution measurements with the VLA (B-, C- and D-array) are aimed at
identifying morphological and kinematical signatures in the \HI caused by
the starburst. Our kinematical data suggest a huge hole in the 
\HI distribution, probably due to the large number of supernovae explosions in 
the center of the galaxy over the past 20~Myr.  
Investigating the large-scale \HI structure, we confirm the existence of a 
possible \HI companion and a so-called \HI bridge east of NGC~1569. 
Furthermore, we report the detection of additional low-intensity \HI halo 
emission, which leads us to suggest a revised halo structure. Based on
the new picture, we discuss the origin of the halo gas and possible 
implications for the evolution of the starburst in NGC~1569.

\end{abstract}


\section{Introduction}

The impact of massive stars on the evolution of galaxies and the 
intergalactic medium is of critical interest, particularly in regard to 
dwarf galaxies.  In low mass galaxies, the combined energy of stellar 
winds and supernova (SN) explosions can lead to the blow-out or 
blow-away of not only the hot metal-enriched gas from the SNe themselves, 
but also a significant fraction of the host galaxy's interstellar medium
\citep{silich01, maclow99, ferrara00}. In the most dramatic cases, starbursts 
in dwarf galaxies have
produced massive outflows of warm and hot ionized gas as seen in 
large-scale H$\alpha$ filaments and X-ray emitting winds 
\citep[e.g.][]{calzetti04, heckman01, summers04}.  Even in the absence 
of large-scale outflows, feedback from massive stars can dramatically 
affect the kinematics and distribution of the ISM in galaxies as suggested 
by the large number of holes in the H~I distributions in nearby dwarf
galaxies such as Holmberg II \citep{puche92}, IC 2574 \citep{walter99}, the 
Magellanic Clouds \citep{staveley97, kim98}, and IC 10 \citep{wilcots98}.

While there is a great deal of interest in understanding the impact of 
starbursts on the ISM and evolution of dwarf galaxies, the question of how 
such starbursts are triggered in the first place remains controversial.
Some investigations find that a merger or interaction with 
gas-rich (dwarf) companions or massive gas clouds may trigger starbursts in 
these objects \citep[e.g.][]{oestlin01,taylor97}. Other studies claim a lack 
of detected companions and suggest stochastic, 
self-propagating star formation or other internal triggers, especially for 
off-center star formation sites and the activity in low-surface brightness 
galaxies \citep[e.g.][]{brosch04}. 

NGC~1569 (UGC~3056, Arp~210, VII~Zw~16) is one of the most actively 
star-forming galaxies in the local universe. This gas-rich dwarf galaxy 
is one of the few nearby dwarfs, in which violent star formation 
is, or has been, taking place. Table~1 
contains some of its basic properties. As a member of the IC~342/Maffei 
group, it is located close to the Galactic plane at a distance of about 
2~Mpc ($2.2 \pm 0.6$~Mpc according to \citet{israel88}, $1.95 \pm 0.2$~Mpc 
according to \citet{makarova03}), hence 1\arcsec = 9.7~pc. NGC~1569 is 
fairly isolated. Except for the dwarf galaxy UGCA~92 at a projected 
distance of 42~kpc there are no near neighbors (Karachentsev, Tikhonov \& 
Sazonova 1994). 

There are several indications that NGC~1569 is in an evolved phase of a 
starburst: it hosts numerous H$\alpha$ filaments, which extend far out into 
the halo (Hunter, Hawley \& Gallagher 1993); it has a distinct cut-off in its 
synchrotron spectrum \citep {israel88b}, extended soft X-ray emission 
and a starburst-driven metal-rich galactic wind (Martin, Kobulnicky \& 
Heckman 2002). Two very 
bright super-star clusters located in the disk \citep[SSCs 
A and B,][]{ables71} appear to be associated with a hole in the 
\HI distribution \citep {israel90} and with a local minimum in the 
emission of the warm molecular gas as traced by the CO line emission
\citep{muehle02}.
Because of its state as a (post-)starburst galaxy, NGC~1569 is an excellent 
laboratory in which to study the impact of massive stars on their 
environment.  Its proximity means that we can study feedback on spatial 
scales of $\sim40$~pc with the resolution available on modern 
interferometers. Here we focus on the distribution and kinematics of 
the \HI in NGC~1569 in order to map the connection between the energy input of 
violent star formation, the outflow of warm and hot gas, and the properties of 
the neutral ISM. We also explore the relationship 
between the previously identified neighboring gas cloud \citep{stil98} 
and the recent star formation history of the galaxy.

The \HI distribution of NGC~1569 was first mapped by \citet{reakes80} and it 
has  a prominent western arm, which extends southward from the disk 
and is located far west of the H$\alpha$ arm. Within the bright \HI ridge 
north of the optical center of the galaxy, \citet{israel90} detected a 
prominent hole, about 100\ pc in diameter, coincident with SSC~A. A 
kinematical study of the \HI emission by \citet{stil02} suggests that 
the gaseous inner disk ($ r < 0.6\ {\rm kpc}$) is quite disturbed. This 
picture is supported by the study of \citet{greve02}, who searched for 
radio supernovae (RSNe) and supernova remnants (SNRs) in the vicinity 
of the super-star clusters and found only a few non-thermal sources in 
the outer regions of an area about 300~pc in diameter around the SSCs.

Our \HI observations, which have an angular resolution higher than those
reported in the past, aimed at scrutinizing the morphology and kinematics 
of the diffuse gas, with an attempt to identify further traces of the 
impact of the past starburst. At the same time, our study of the 
large-scale \HI distribution and kinematics has uncovered possible evidence 
for a 
trigger of this burst. Sect.~2 describes the observations and data reduction, 
also addressing the problem of the Galactic foreground emission and the issue 
of missing flux. In Sect.~3, the detailed \HI distribution and kinematics of 
the disk of NGC~1569 are discussed and compared with the available \Ha 
morphology and velocity information. In 
Sect.~4, we present the faint \HI halo structure around NGC~1569 and discuss 
its possible origin and the resulting implications for the starburst. Sect.~5 
summarizes our results.

\section{Observations and Data Reduction}

The \HI observations were carried out with the Very Large Array\footnote{The 
VLA is a facility of the National Radio Astronomy Observatory (NRAO), itself 
a facility of the National Science Foundation operated under cooperative 
agreement by Associated Universities, Inc.} (VLA) in the B-configuration 
(14 hours of observing time in April 1993), in the C-configuration (7 hours in 
June 1993), and in the D-configuration (3 hours in September 1992). 
We chose a correlator bandwidth of 1.56\,MHz and applied
Hanning smoothing online, which resulted in a data cube of 127 channels 
with a velocity resolution of $v=2.6\ {\rm km\,s^{-1}}$. The data were 
calibrated with the standard phase calibrator 0614+607 and the flux calibrator 
0137+331. We then subtracted the continuum emission in the $u$--$v$ plane 
using the emission-free channels on both ends of the spectrum not affected by 
Galactic foreground emission.

For the deconvolution and imaging of our data, the standard AIPS task IMAGR 
was applied. We tried several weighting schemes by varying the weighting
parameter ROBUST \citep{briggs95}, thus producing image cubes with about 
natural weighting (ROBUST = 5), or with a scheme between uniform and natural 
weighting (ROBUST = $0 \ldots 3$). From these cubes, we chose two data sets 
for our analysis: a high-resolution cube (ROBUST = 1) with a beam width of 
$7\farcs5 \times 7\farcs3$ FWHM and an rms noise level of 
$0.53\ {\rm mJy\,beam^{-1}}$ in the channel maps corresponding to an HI 
column density of $N_{\rm HI} =2.74 \cdot 10^{19}\ {\rm cm^{-2}}$, which   
emphasizes the small-scale structure in the disk of NGC~1569, 
and a low-resolution cube 
(ROBUST = 5) with a beam width of $13\farcs7 \times 13\farcs1$ FWHM and a 
noise level of $0.53\ {\rm mJy\,beam^{-1}}$ corresponding to 
$N_{\rm HI} =0.84 \cdot 10^{19}\ {\rm cm^{-2}}$, which is sensitive to the 
large-scale low-column density features. Each cube was cleaned down to 
the noise level. 

The filtering nature of the interferometer works to our advantage: by 
combining VLA observations 
taken in the B-, C-, and D-configuration, our data include structures 
from $\sim 4\arcsec$ (B-array) to $\sim 15\arcmin$ (D-array). Thus, all 
of the large-scale structure except for possibly existing extended halo 
features physically connected with NGC~1569 will be present in the 
interferometric data, while much of the unwanted large-scale Galactic 
foreground structure is efficiently filtered out. A caveat 
remains, in the sense that if there are small-scale Galactic features in 
the line of sight toward the dwarf galaxy with velocities similar to 
those of NGC~1569, they cannot be separated from the structures of the 
dwarf galaxy.

We show every other channel map, starting at $v=-10.2$~km\,s$^{-1}$ in 
Figure~\ref{chanuni}.  
The channel maps of the high-resolution cube clearly show Galactic 
foreground emission in the velocity range from $v=+5\ {\rm km\,s^{-1}}$ 
to $v=-60\ {\rm km\,s^{-1}}$. Two features of Galactic foreground emission 
seem to be localized in a broad stripe north of NGC~1569, extending from the 
southeast to the northwest ($v=-39 \ldots -61\ {\rm km\,s^{-1}}$), and in a 
region at the southern rim of the channel maps ($v=-39 \ldots -59\ {\rm 
km\,s^{-1}}$), respectively. These could be easily removed from the data cube. 
The emission in the velocity range $v > -26\ {\rm 
km\,s^{-1}}$ changes almost randomly from one map to the next with some 
structures persisting over several channels and is clearly foreground emission.
At velocities $-18$~\kms\ $> 
v > -30$~\kms , the line of sight toward NGC~1569 shows \HI emission, only 
part of which is likely to be extragalactic. As we cannot determine the 
amount of the \HI emission belonging to the dwarf galaxy, we have to treat 
these channels with caution in our quantitative analysis. 

The higher sensitivity of the low-resolution cube reveals even more Galactic 
foreground emission with the same general structure in the corresponding 
channel maps (not shown here). 
Luckily, much of the foreground emission is spatially separated from 
NGC~1569 in most channels. Nevertheless, it complicates the analysis of 
the spectrum of the galaxy, because at some 
velocities ($v = +5 \ldots -28$~\kms\ and $v = -44 \ldots -54$~\kms),
the position of NGC~1569 is coincident 
with a negative side lobe of the large-scale foreground emission. 
As a means to largely remove the Galactic structure while keeping the 
large-scale emission of NGC~1569, we blanked the innermost pixels in the 
Fourier plane of the affected channel maps. As a result, the negative trough 
has been nearly completely removed from the data cube.

In order to derive the overall \HI spectrum of NGC~1569, we applied this
blanking to the low-resolution cube, blanked the noise in the channels at the 
3~rms level and removed the Galactic 
foreground emission that is spatially separated from the dwarf galaxy. The 
slight asymmetry that we see in the resulting spectrum of the \HI emission in 
NGC~1569 (Fig.~\ref{vlaspec}) may be due to Galactic foreground emission 
that is in the line of sight toward NGC~1569 and/or may reflect a true 
asymmetry in the velocity field.

A standard issue in the analysis of interferometric data is possibly
missing flux due to the reduced sensitivity of any interferometer to
large-scale structure. We used Effelsberg observations to also obtain a good
single dish measurement of NGC1569 \citep{muehle03}. The total HI flux derived 
from these data is consistent with that in the multi-configuration VLA data. 
Thus, a combination of the interferometric data with the single dish spectrum 
was not necessary. The total \HI mass of NGC~1569 resulting from the VLA data 
is $M_{\rm HI}= 7 \cdot 10^7\ {\rm M_{\odot}}$.


\section{The \HI Disk} \label{disk}

For a comprehensive study of the small-scale structure of the main body of 
NGC~1569, which we call the disk, we made channel maps (Fig.~\ref{chanuni})
from the original high-resolution cube. We also produced a map of the \HI 
column density from the high-resolution data cube, which we corrected for 
the Galactic foreground emission, excluding the channels dominated by 
Galactic foreground emission and blanking the noise at the 3~rms level 
(Fig.~\ref{mom0}). The maps of the velocity field (Fig.~\ref{mom1}) and 
of the velocity dispersion (Fig.~\ref{mom2}) have been calculated in the 
same manner.

Accounting for
precession from the B1950 to the J2000 coordinate system, the position angle 
for the major axis determined from the column density map is in very good 
agreement with that determined by \citet{reakes80}. 
In order to examine the data cube for kinematical peculiarities, we rotated 
the uncorrected high-resolution cube clockwise by 23\degr, so that 
the major axis of the \HI disk is aligned with the x-axis. From this data 
cube, we produced 
position-velocity maps of the central region, with a separation of about one 
beam width in declination (Fig.~\ref{pv1}). Interpreting the $p-v$ maps, 
one has to keep in mind that the upper part of the maps, especially at $v 
> -20$~\kms , may be affected by the Galactic \HI foreground. The main 
features of the \HI distribution and of the \HI kinematics of the disk are 
summarized in Table~2.

The global velocity field of NGC~1569 (Fig.~\ref{mom1}) is consistent with 
slow solid-body rotation, with the eastern side of the gaseous disk receding 
at radial velocities of up to $0\ {\rm km\,s^{-1}}$ and the western side 
approaching at velocities $v > -150\ {\rm km\,s^{-1}}$. However, a 
closer look at the velocity field (Fig.~\ref{mom1}) and the $p-v$ maps 
(Fig.~\ref{pv1}) shows that in the central 900~pc of the disk, the velocity 
distribution is broader and seems to be independent 
of the position along the major axis. The decoupling of this region is in 
excellent agreement with the results of \citet{stil02}, who state that the 
center of the gaseous disk of NGC~1569 ''is strongly disrupted'' 
(cf.\ Section~\ref{holes}). 

South of the disk, we find a roughly circularly shaped region of \HI emission, 
the ``southern blob'' (Fig.~\ref{mom0}), which seems to be connected to the 
disk in the southeast by an \HI bridge, the ``junction''. According
to the position-velocity maps (Fig.~\ref{pv1}), the southern blob consists of 
two velocity components. The main part of the blob has a velocity distribution 
in the range of \mbox{$v = -100 \ldots -50$~\kms,} while the protrusion, which 
may or may not be physically associated with the southern blob, shows 
velocities of $v > -50$~\kms . It is worth to note that the 
region in the disk opposite to this area shows a high radial velocity of 
$v > -60\ {\rm km\,s^{-1}}$, too.  These features seem to be spatially 
associated with \Ha filaments (Section~\ref{comparison})

The main body is elongated to the west, ending in an \HI arm which points 
westward. With the better sensitivity of the low-resolution cube, we can 
trace the \HI arm far out into the halo. Thus, we treat it as a halo feature, 
which will be further discussed in Sect.~\ref{halo}.


\subsection{Holes in the \HI Disk: Swiss Cheese or Donut?}
\label{holes}

Using the high-resolution data cube, we searched for holes by analyzing the 
individual channel maps and the position-velocity diagrams. 
In order to minimize the unavoidable personal bias in the detection of holes, 
each author has compiled a list of candidate holes. We used the standard 
criteria for holes as defined by \citet{brinks86}: {\em 1)} the center of 
the hole must be stationary in the maps in which it is seen in order to 
avoid mixing up kinematical effects with holes; {\em 2)} the hole must have 
a good contrast with respect to its surroundings in all relevant channel maps; 
and {\em 3)} the shape of the hole must be clearly defined and adequately 
described by an ellipse. Because neighboring pixels in our data are 
independent, we relaxed Brinks' and Bajaja's fourth requirement that a hole 
should be visible in at least three successive channel maps.
Except for the prominent \HI chimney and one very tentative detection of a 
hole signature at  $\alpha_{2000}=04\h30\m42\s$, 
$\delta_{2000}=64\degr51\arcmin06\arcsec$, 
where the high-resolution \HI column density map (Fig.~\ref{mom0}) shows a 
local minimum, we have found no holes that satisfy the search criteria.

Several factors may impede the detection of \HI holes in NGC~1569: 
First, the maximum expansion velocities of the shells surrounding the 
holes are naturally found along the steepest density gradient. In the case of 
NGC~1569, the axis perpendicular to the disk is at an angle of $63\degr$ from 
our line of sight, which may make the dominant velocity component difficult to 
detect. As the total column density of an individual pixel is the sum of the 
emission from a large volume along the line of sight, homogeneous regions 
along a line of sight could easily mask the depression in the \HI emission 
characterizing an \HI hole in the column density map. Note however, that 
numerous holes have been detected in IC~2574, a dwarf galaxy with an even 
higher inclination and a scale height of 350\,pc \citep{walter99}.
Second, the kinematical study by \citet{stil02} indicates that the 
center of the gaseous disk of NGC~1569 is quite disturbed. Strong 
turbulence may mask the signature of holes in the position-velocity maps, but 
may also destroy these features.
Third, we note that some channel maps at high velocities are 
so contaminated by Galactic foreground emission that it is impossible to 
determine the extragalactic \HI flux and morphology. Any possible \HI shells 
with velocities $v>-20$~\kms \ can not be detected in our data. \\

The complete lack of detectable holes, with the notable exception of the
\HI chimney at the position of the two SSCs, may reflect an actual absence of 
large holes in (the center of) the disk of NGC~1569. The destruction of a 
possible past 
hole-dominated \HI structure (``Swiss cheese'') may have been caused by gas 
accretion (Section \ref{halo}) and the extraordinary strength of the starburst 
itself. We have noted earlier (Section~\ref{disk}) that the characteristic 
solid body rotation seems to commence at a radius of about 450~pc, while the 
gas kinematics in the center of the disk is consistent with no systematic 
rotation. This velocity structure may indicate that the low-angular momentum 
gas in the center has been blown to larger radii. Taking into account the 
observed biconical outflow along the minor axis of the galaxy, the remnant of 
the swept-up gas may be located in a ring in the disk (``donut''). Such a 
scenario is in good agreement with the high-resolution \HI column density and 
velocity maps. Due to the high inclination of the disk, the 
\HI emission from the nearer and the farther side of the ring most likely 
overlap, resulting in a broad high-column density ridge. The \HI chimney is 
then located in the southern part of the ring, while the more homogeneous \HI 
distribution in the northern part gives rise to the higher column density 
found in the projection north of the chimney (Fig.~\ref{mom0}). Adopting the 
3-D model of \citet{martin02} and assuming that the ring lies in the plane of
the major axis of the galaxy, the northern part of the ring is closer to the 
observer than the southern one. The difference in the average velocity 
($\sim 15$~\kms, Fig.~\ref{mom1}) between the nearer (northern) and the 
farther (southern) side of the ring might trace the remnant expansion of the 
hot central bubble, before it bursted out of the disk.

With a supernova rate of $(5 \ldots 25) \cdot 10^3$~SNe\,Myr$^{-1}$ 
over the last $\sim30$~Myr \citep{vallenari96,waller91}, the 
starburst has been intense enough to sweep the whole central region from 
neutral atomic hydrogen, heat the dust and the molecular gas to 
unusually high temperatures \citep{lisenfeld02, muehle02}, and to form a 
strong galactic wind \citep{martin02}. Other evidence for a depletion of 
neutral gas in the center 
of the galaxy may be the paucity of radio supernovae and supernova remnants 
\citep{greve02}. \citet{anders04} report that the younger star clusters have a 
significantly lower mass compared to the older ones 
and suggest that the strong radiation field and the burst of supernova 
explosions prevented the assembly of large molecular clouds in the observed 
central region of NGC~1569.\\ 

The formation of a donut \HI structure as a consequence of violent 
star formation may indeed occur frequently in dwarf galaxies. NGC~4449, 
another star-forming dwarf galaxy, is conspicuous because of the lack of \HI 
holes in its inner disk (Hunter, van Woerden \& Gallagher 1999). Instead, the 
\HI distribution is dominated 
by a huge ring. The dwarf galaxy Holmberg I shows a ring-like \HI morphology, 
too, with a pronounced depletion of the \HI column density in the center of 
the galaxy and evidence for a blow-out \citep{ott01}. 
In a recent sample of late-type dwarf galaxies, one third of the 
galaxies are characterized by a high-column density feature with 
$N_{\rm HI} > 10^{21}$~cm$^{-1}$ in the form of a ring and low \HI column 
densities in the center \citep{stil02a}. The rings, which appear to be broken 
in some cases, usually have a size comparable to the underlying \HI disk.
With NGC~1569, we may have found a prototype of this interesting class of 
galaxies viewed at a large inclination.


\subsection{Comparison of \HI and H$\alpha$}
\label{comparison}

\citet{hunter93} found three shells and numerous filaments in the \Ha 
distribution (Fig.~\ref{ha}). Two adjacent shells are located at the eastern 
rim of the \HI 
ridge south of the center of the eastern ridge. The third one is located 
south of the \HI chimney and the \Ha trough.
In order to confirm possible associations between the \Ha morphology and the 
\HI distribution, it is necessary to obtain velocity information for the 
different \Ha features. Tomita, Ohta, \& Saito (1994) performed long-slit 
\Ha spectroscopy covering the optical extent of NGC~1569 with 11 slits of 
5\farcm0 in length and 1\farcs8 wide. Fitting one- and two-component 
Gaussians to the data, they produced position-velocity diagrams, which 
can be compared with our \HI data. The analyzed slit positions are shown in 
Figure~\ref{slits}, the corresponding \HI $p-v$ maps in Figure\ 
\ref{slitmaps}. \citet{tomita94} found several ``small splits'' (SS) in 
the \Ha line profiles, which they interpret as expanding \Ha bubbles.

The strongest \Ha emission is located south of the major axis of the \HI 
distribution, suggesting that most of the \HII regions may be located in the 
southern part of the \HI ring (Fig.~\ref{ha}). At the same position, 
\citet{tomita94} found the kinematical signature of a shell (SS~1), whose 
velocities are consistent with the corresponding \HI velocity distribution.
The pronounced chimney in the \HI ridge coincides with the position of the 
super-star clusters and with a minimum in the \Ha emission (Fig.~\ref{ha}, 
left). This may indicate that the SSCs
have blown away the ISM in their surroundings. Shock waves 
from (multiple) supernova explosions may have ionized the bubble and 
triggered new star formation in the swept-up shell, resulting in today's \HII 
regions around the SSCs. Neutral atomic hydrogen can still exist outside of 
this area. 

The only clear spatial correspondence between the \Ha filaments and the \HI 
structure occurs for filaments \#7 and \#8 
\citep[in][Fig.~\ref{ha}, right]{hunter93}, which delineate the northern 
and the southeastern rim of the southern blob. These features, combined with 
the main disk and the prominent \Ha arm \citep[filament \#6 in][]{hunter93} 
may be the remnant of a huge \Ha shell. The kinematical information of 
\citet{tomita94} allow for an association between the tip of the \Ha arm and 
the high-velocity \HI gas south of the disk.  

At the position of the \Ha shell SS~5 in slits J and K 
($\alpha_{1950} = 04\h 26 \m 06\s \ldots 09\s$, 
$\delta_{1950} = 64\degr 43\arcmin 52\arcsec \ldots 58\arcsec$), a local 
minimum in the \HI emission can be found in the corresponding $p-v$ maps, most 
noticeably in the \HI slit J, which reaches velocities of 
up to $-80$~\kms. This is consistent with the higher velocity component in \Ha 
of about $-70$~\kms. The lower \Ha velocity component of about $-125$~\kms \ 
has no counterpart in the \HI gas. 

We did not identify \HI features corresponding to the other H$\alpha$ shells 
\citep{tomita94}. This may be due to the fact that the Gaussian 
fit has compressed a broad distribution of \Ha velocities to one or two values.
Another possible explanation is that the distribution of the neutral gas may 
be more strongly affected by the turbulence in the inner disk than the \Ha 
morphology, resulting in a higher velocity dispersion smearing out the 
kinematical signatures of a shell.


\section{The Halo Structure} \label{halo}

Utilizing the high sensitivity of the low-resolution \HI cube, we 
investigated the gaseous halo of NGC~1569. \citet{stil98} reported the 
detection of extragalactic low-intensity \HI emission close to NGC~1569, 
which they named NGC~1569-HI and which they suggested to consider as a 
companion to the dwarf galaxy. The existence and the velocity structure 
of this feature and of the so-called \HI bridge, which seems to connect 
NGC~1569-HI and the disk, are readily confirmed in our \HI column density 
map and velocity field derived from the naturally weighted and 
foreground-corrected VLA data cube (Fig.~\ref{stream.mom0} and 
\ref{stream.mom1}). In addition, we have found very faint filamentary 
\HI emission, the``\HI stream'', south of the disk of NGC~1569, which 
appears to stretch from the western end of the \HI bridge east of the 
galaxy, to the western \HI arm. There is also faint emission northeast 
of NGC~1569-HI, which we call the ``northern blob''. The general 
properties of the detected condensations (Fig.~\ref{streamoverlay}) 
and their \HI masses are summarized in Table~3. The corresponding spectra 
are shown in Fig.~\ref{halospec}. In contrast to the other features, the 
northern blob seems to be a superposition of many small 
\HI clouds with slightly different velocities, and is possibly of Galactic 
origin.

\subsection{The \HI arm}

It is remarkable that the \HI arm can be seen in the halo as well as in the 
disk, reaching inward all the way into the main body of NGC~1569.
The position of the inner \HI arm suggests that it might be an extension of 
the apparent \HI high column density ridge. 
But the velocity field (Fig.~\ref{mom1}) and the $p-v$ maps (Fig.~\ref{pv1}) 
imply that this structure forms a different kinematical component: 
The western end recedes at $v \approx +8\ {\rm km\,s^{-1}}$, while the 
eastern side approaches the observer at a speed of about $-13$~km\,s$^{-1}$ 
relative to the systemic velocity of $v_{\rm sys}= -80$~\kms. The red-shifted 
velocity component at ${\rm offset} = -40\arcsec$ visible in the southwestern 
$p-v$ maps 
might be the root of the \HI arm. The entire arm including the region where it 
connects to the inner disk is well west of the optical disk and the \Ha and 
X-ray halos \citep{hunter93, martin02}. 
Probably related to the interface between disk and arm, the radial velocity 
dispersion in the southern part of the western ridge is twice as high as the 
average velocity dispersion of $\sigma_{\rm v} \approx 15\ {\rm km\,s^{-1}}$ 
(Fig.~\ref{mom2}). This ``hot spot'' (Table 2) may indicate a region of 
velocity crowding, i.e.\ the superposition of several kinematical components.
Note that the root of the \Ha arm coincides with the hot spot, while the tip 
of the arm seems to be spatially and kinematically associated with the 
high-velocity region in the southern \HI disk \citep[cf.][]{tomita94}. Thus, 
we may see here a superposition of the kinematical signature of the disk, the 
peculiar velocity of an extension of the \HI arm, and possibly a contribution 
of \HI gas associated with the \Ha arm.

\subsection{The \HI stream}

Although parts of the \HI stream have a velocity range that is very close 
to that typical for the Galactic clouds in this region, there are several 
indications that the detected \HI clouds are not a condensation in the Galactic
 foreground, 
but rather belong to the halo of NGC~1569. A first hint at their extragalactic 
nature is the conspicuous structure, which seems to stretch from the western 
end of the \HI bridge to the \HI arm in the west (Fig.~\ref{stream.mom0}). 
Their filamentary appearance does not resemble any of the Galactic 
condensations that we find in the channel maps. The velocity within the 
stream continuously shifts from about $-70$~\kms \ near the \HI bridge to 
a maximum of $-64$~\kms \ in the stream, and then back to about $-79$~\kms\ at 
the end of the \HI arm (Fig.~\ref{stream.mom1}). As part of an extended 
halo structure, the puzzling kinematical behavior of the \HI arm would find a 
compelling explanation. The halo feature may be \HI gas that has been 
pulled out of the disk by tidal forces due to an interaction with a nearby 
galaxy or a condensation in the IGM that had come 
close to NGC~1569 and has then been tidally disrupted in its gravitational 
potential. Thus, the coherent morphology and velocity field of the structure 
suggest that the \HI arm, the \HI stream, the bridge and the 
companion are physically associated, and we consider it unlikely that the 
detected clouds are part of the Galactic foreground.\\

\subsection{Possible causes of the halo structure}

With a projected distance of only 42~kpc, UGCA~92 is the closest neighbor to 
NGC~1569. It is located at about the same distance ($\sim$ 1.8~Mpc), and also
has a comparable radial velocity \citep[$-99$~\kms,][]{karach94}. Is it 
conceivable that UGCA~92 has pulled a tidal arm out of the disk of NGC~1569 
in the course of a close encounter, corresponding to the observed halo 
structure? Simple estimates suggest that for the tidal force to exceed the 
internal gravitational force in an encounter without physical contact, the 
disturbing galaxy should be more massive than the disturbed galaxy: 
$$ M_{\rm comp} \geq M \left( \frac{l}{R} \right)^3 $$
with $M$ and $R$ the mass and the radius of the considered galaxy, 
$M_{\rm comp}$ the mass of the companion galaxy and $l$ the distance between 
the two galaxies \citep{saslaw85, campos93}. 
UGCA~92, though, is a low-surface brightness galaxy with less \HI mass 
than NGC~1569 \citep{devau91}.   
N-body simulations of the interaction between the dwarf galaxies NGC~4449 and 
DDO~125 suggest that tidal structures can be seen in the first 
few $10^8$ years following such an encounter \citep{kohle99}. Given the 
low mass of UGCA~92, the encounter would have to have been very close in order 
to make tidal forces a significant factor, thus requiring a favorable 
geometry and a relative velocity of about 100~\kms\ or more in 
the plane of the sky for such an encounter to have occurred about $10^8$~years 
ago. A possible 
encounter should have disturbed the \HI distribution of UGCA~92 even more than 
that of NGC~1569. Unfortunately, we did not find any high-resolution \HI 
image of UGCA~92. 
Thus, at this stage of our investigation, we can not completely rule out a 
possible interaction between NGC~1569 and UGCA~92 in the past, but we consider 
it unlikely to be the cause for the detected \HI gas in the halo.\\

The observed halo structure may also be interpreted as the 
remnant of an infalling \HI cloud. In the past, an \HI cloud may have 
come close to the gravitational potential of NGC~1569 and been 
tidally disrupted while falling toward the disk of the dwarf galaxy. 
There are no traces of any internal rotation in the halo structure, which 
implies that the infalling cloud would have been nearly face-on or that 
tidal forces would have completely disrupted the velocity field of 
the cloud. 

A number of observations support the scenario of an infalling \HI cloud as the 
origin of the halo structure. To begin with, the existence of an \HI companion 
close to a star forming dwarf galaxy is not unusual. 
In a high-resolution \HI survey of a volume-limited sample of 21 
blue compact dwarf galaxies one or more 
companions have been detected near 12 galaxies \citep{taylor95, taylor96}. 
Four of the companions have no optical counterparts and 
may thus be \HI clouds. In a similar study, \citet{wilcots96} found \HI 
clouds with masses of $\sim10^7M_{\odot}$ around four out of five barred 
Magellanic-type spiral galaxies of their sample. Most of the clouds seem to be 
interacting with their primary galaxy. Recently \HI clouds have been
detected in the halo of M~31, too, which are likely similar
to the high-velocity clouds of the Milky Way and which may be debris from 
recent mergers or interactions \citep{thilker03}.
An example for a tidally disrupted \HI 
cloud in the process of being accreted onto the disk of a spiral galaxy has 
been presented by \citet{phookun93}: An extended stream of small 
\HI clouds seems to wrap around the grand-design spiral galaxy NGC~4254 in a 
spiral pattern that indicates that this gas will probably merge with the 
primary galaxy in the near future and that other gas may have already been 
accreted. The \HI mass of these clouds is 
$\sim2.3 \cdot 10^8 M_{\odot}$. The halo structure of NGC~1569 is very similar 
to the disrupted \HI cloud around NGC~4254 appearing like a scaled-down, 
tilted version of the latter's \HI stream. This suggests accretion processes 
that are in a similar phase.  
\citet{wilcots98} found that the process of galaxy formation by the 
accretion of gas from a primordial reservoir may still go on in the dwarf 
galaxy IC~10. However, there is no indication for such a huge gas halo around 
NGC~1569. Thus, any infalling \HI gas is probably an intergalactic cloud that 
survived in this sparsely populated region of the universe until the encounter 
with NGC~1569.

More evidence in favor of the scenario of an infalling cloud can be found in 
NGC~1569 itself. It is conspicuous that the X-ray emission, the \Ha filaments 
and the radio continuum distribution all extend over about the same region, 
well inside the area enveloped by the \HI halo, and in particular with a sharp 
boundary east of the \HI arm. Here, at the location of the prominent \Ha arm, 
\citet{martin02} found evidence for shocked, X-ray emitting gas, while our
kinematical study of the \HI emission reveals the superposition of two or 
three 
velocity components at the hot spot close to the arm. Thus, it is tempting to 
assume that this region is not only the remnant of a huge shell 
(Sect.~\ref{comparison}), but also the impact zone of the gas falling onto 
the disk of NGC~1569. The emergence of the starburst and its extraordinarily 
high star formation rate could then be explained by the accretion of \HI gas 
with a mass of about $10^7M_{\odot}$, which replenished the gas reservoir of 
NGC~1569.
In this picture, the halo structure observed today would be only the remnant 
of an intergalactic cloud of originally a few times $10^7M_{\odot}$. Note that 
this mass estimate is in very good agreement with the masses of the other 
mentioned streams, but also only about an order of magnitude lower than the 
estimated total mass of NGC~1569.


\section{Conclusions} \label{summary}

We observed the detailed \HI emission of the post-starburst dwarf galaxy 
NGC~1569 with the VLA (B-, C-, and D-configuration). The Galactic 
foreground \HI emission was minimized by spatial frequency filtering and 
subsequent careful blanking of still affected regions. A comparison of the VLA 
data with a single-dish spectrum indicates little or no missing flux in the 
interferometric data cube. Our main results are:

 The \HI line spectrum is centered at $v_{\rm sys} \approx -80$~\kms\ and 
suggests a total \HI mass of $M_{\rm HI} = 7.0 \cdot 10^7\, M_{\odot}$.

 In the central region of the disk ($r< 450$~pc), the kinematics appears 
to be highly disturbed, whereas the outer regions show a slow solid body 
rotation. We suggest that the gas in the center of the galaxy, which is 
characterized by low angular momentum, may have been blown to larger radii 
by the strong stellar winds and multiple supernova explosions that also drive 
the galactic outflow, ultimately resulting in a high-density \HI ring (donut) 
and a depletion of \HI gas in the center of NGC~1569.

 A pronounced chimney in the \HI distribution, which probably lies in the 
southern, remote part of the ring, is coincident with the position 
of the two prominent super-star clusters and with a double-peaked \Ha line 
profile. 
Our interpretation  is that strong stellar winds from massive stars and 
supernova explosions in these clusters may have blown away the surrounding 
ISM and may have led to secondary star formation in the swept-up material,
now seen as a ring of \HII regions.

 The prominent western \Ha arm may be associated 
with some of the \HI emission south of the disk. Together with two southern 
\Ha filaments and the southern \HI blob, this structure might be the remnant 
of a huge shell.  

 We confirm the existence and the velocity structure of \HI gas in the halo 
of NGC~1569, apparently consisting of the companion ``NGC~1569-HI'' and an 
\HI bridge. In our high-sensitivity data, we found additional faint \HI 
emission in the southern halo of NGC~1569. 

 The \HI arm of NGC~1569, an apparent extension of the disk to the west, has 
a kinematical behavior that differs strongly from the velocity field in the 
disk, but seems to be continued in the newly detected halo gas. Its peculiar 
velocity can be traced far into the western part of the disk. The hot spot, a 
region of \HI velocity crowding in the western disk of NGC~1569 can thus be 
interpreted as the superposition of the kinematical components of the \HI disk,
 the \HI arm, and possibly an \HI counterpart of the \Ha arm. 

As a new comprehensive picture of the halo around NGC~1569, we propose that 
the known halo features, the newly detected \HI gas in the southern halo 
and the western \HI arm form a coherent halo structure, which can be traced 
from a location in the center of the disk via the \HI arm, the stream and the 
bridge to NGC~1569-HI. 
Thus, a continuous halo feature seems to wrap around the disk of NGC~1569 
from the companion east of the galaxy to the western \HI arm and maybe even 
further. 

 The most likely causes for the detected \HI halo gas are tidal interaction
with a nearby galaxy or accretion of neutral atomic hydrogen. We suggest that 
the halo gas may be the remnant of an intergalactic \HI cloud that has 
been tidally disrupted when 
approaching the dwarf galaxy and may already have been partially accreted onto 
the disk of the galaxy. Such an event would have likely triggered a starburst.
The unusually high star formation rate in NGC~1569 could thus be explained by 
the 
replenishment of the \HI reservoir in its disk by the newly accreted material.



\acknowledgments

We are grateful to C.\ Br\"uns for performing the observations at Effelsberg 
and to D.\ Hunter for sending us her digital \Ha image. We thank Ernie 
Seaquist, Dave Westphal 
and Bill Waller for helpful comments. S.M.\ acknowledges the Deutsche 
Forschungsgemeinschaft for the award of a fellowship of the Graduiertenkolleg 
"The Magellanic System, Galaxy Interaction, and the Evolution of Dwarf 
Galaxies", and support under grant SFB~494. E.M.W.\ thanks the 
Graduiertenkolleg "The Magellanic System, Galaxy Interaction, and the 
Evolution of Dwarf Galaxies" for their hospitality in support of this 
project. This work was partially supported by the National Science 
Foundation through grant AST 0098438 to E.M.W. This research has made use 
of the NASA/IPAC Extragalactic Database (NED) which is operated by the Jet 
Propulsion Laboratory, California Institute of Technology, under contract 
with the National Aeronautics and Space Administration, and NASA's 
Astrophysical Data System Abstract Service (ADS).\\





\clearpage


\begin{figure}
\epsscale{0.8}
 \caption{Every alternate channel map of the original VLA high-resolution 
 \HI cube \mbox{($7\farcs5 \times 7\farcs3$)}.
The contour lines in the individual channels denote levels of 3~rms and 10~rms.
The first channels are clearly contaminated by Galactic foreground emission. 
The localized Galactic foreground structures lie outside of the plotted area.
 \label{chanuni}}
\end{figure}

\begin{figure}
\figurenum{1b}
\epsscale{0.8}
\caption{ }
\end{figure}

\begin{figure}
\figurenum{1c}
\epsscale{0.8}
\caption{ }
\end{figure}


\begin{figure}
\caption{\HI spectrum of NGC~1569 derived from the foreground-corrected 
        low-resolution cube ($13\farcs7 \times 13\farcs1$). The 
        noise has been blanked at the 3~rms 
        level in the channel maps. 
\label{vlaspec}}
\end{figure}

\begin{figure}
 \caption{\HI column density map of NGC~1569 calculated from the channels 
  at $v=-144 \ldots -23$~\kms\ of the high-resolution data cube 
  ($7\farcs5 \times 7\farcs3$), which has been corrected for the large-scale 
  Galactic foreground emission. The contour lines denote the column density 
  levels $N_{\rm HI}=0.05,\ 0.5,\ 2.0,\ 4.0,$ and 
  $6.0 \cdot 10^{21}\ {\rm cm^{-2}}$. A pronounced chimney splits the 
  high-column density ridge into an eastern and a western part. 
 \label{mom0}}
\end{figure}

\begin{figure}
 \caption{Velocity field of NGC~1569 calculated from the channels at 
  $v=-144 \ldots -23$~\kms\ of the foreground-corrected, high-resolution 
  data cube. The contours outline the velocities $v = -120 \ldots$ $-40$~\kms\ 
  in steps of 10~\kms. 
 \label{mom1}}
\end{figure}

\begin{figure}
 \caption{Distribution of the velocity dispersion 
  ($\sigma_{\rm v}= 0 \ldots 31$~\kms) of the foreground-corrected 
  high-resolution data cube in NGC~1569. The contours are the same as in 
  Fig.~\ref{mom0}. 
 \label{mom2}}
\end{figure}

\begin{figure}
\epsscale{0.8}
\caption{Position-velocity maps of the uncorrected, rotated high-resolution 
 cube. For an easy interpretation of the kinematical features, we rotated 
 the original cube such that the major axis of the disk is aligned with the 
  x-axis. The zero offset position and the location of the central plane are 
  shown in Fig.~\ref{mom0}. The label of each diagram denotes the distance 
  along the minor axis from the central plane. 
 \label{pv1}}
\end{figure}

\begin{figure}
\figurenum{6b}
\epsscale{0.8}
\caption{}
\end{figure}

\begin{figure}
\caption{Two representations of the \Ha image of NGC~1569 kindly provided by 
 \citet{hunter93}. Three holes and numerous filaments have been identified. 
  As in Figure~\ref{mom0}, the contours denote the column density levels 
  $N_{\rm HI} =0.05,\ 0.5,\ 2.0,\ 4.0,$ and 
  $6.0 \cdot 10^{21}\ {\rm cm^{-2}}$.
  \label{ha}}
\end{figure}

\begin{figure}
\caption{Analyzed slit positions of the long-slit \Ha spectroscopy by 
 \citet{tomita94}. The greyscale image shows the \Ha emission, the contour 
 lines denote the column density levels $N_{\rm HI}=0.05,\ 0.5,\ 2.0,\ 4.0,$ 
 and $6.0 \cdot 10^{21}\ {\rm cm^{-2}}$ of our foreground-corrected 
 high-resolution map (Fig.~\ref{mom0}). Note that the coordinates are in 
 B1950, such as to facilitate the comparison with the diagrams by 
 \citet{tomita94}.
\label{slits}}
\end{figure}

\begin{figure}
\caption{These \HI position-velocity maps correspond to the diagrams of the 
 \Ha long-slit spectroscopy by \citet{tomita94}. Like the slits, the maps have 
 a separation of $6\farcs0$ in declination with slit A at 
 $\delta_{1950}=64\degr 44\arcmin 52\arcsec$. The offset in right ascension is 
 in arcseconds measured from the eastern end of the slits at 
 $\alpha_{1950}=04\h 26\m 18\fs3$. The width of the slits of $1\farcs8$ is 
 matched by a pixel size of $2\farcs0$ in the \HI maps.
\label{slitmaps}}
\end{figure}

\begin{figure}
\caption{\HI column density map of the halo derived from the 
 foreground-corrected low-resolution cube ($13\farcs7 \times 13\farcs1$, 
 $v=+3 \ldots -214$~\kms). The noise has been 
 blanked at the 2~rms level in the channel maps. 
 NGC~1569-HI is the feature at $\alpha_{2000}=04\h 32\fm2$, 
 $\delta_{2000}=64\degr 49\arcmin$.
 \label{stream.mom0}}
\end{figure}

\begin{figure}
\caption{The velocity field of the condensations in the halo derived from 
 the foreground-corrected low-resolution cube ($13\farcs7 \times 13\farcs1$, 
 $v=+3 \ldots -214$~\kms). 
 \label{stream.mom1}}
\end{figure}

\begin{figure}
\caption{\HI column density map of the halo derived from the 
 foreground-corrected low-resolution cube. The 
 rectangles show the groups of clouds in the halo which were averaged in 
 the spectra of Fig.~\ref{halospec}. The corresponding \HI masses are listed 
 in  Table~3. \label{streamoverlay}}
\end{figure}

\begin{figure}
\caption{Spectra of small groups of \HI clouds in the halo (cf.\  
 Fig.~\ref{streamoverlay}). In each channel of the data cube, the areas 
 dominated by Galactic  
 foreground \HI emission have been blanked. Thus, some of the spectra (B3, 
 S1 ... S4, A) may be truncated. The dip to negative values in the baseline 
 of the S1 spectrum is due to the strong negative side lobe of NGC~1569 at 
 velocities of $v=-80 \ldots -150$~\kms. 
 \label{halospec}}
\end{figure}

\begin{figure}
\figurenum{13b}
\caption{}
\end{figure}

\begin{figure}
\figurenum{13c}
\caption{}
\end{figure}


\clearpage

\begin{table}
\begin{center}
 \caption{Basic Properties of NGC~1569}
\begin{tabular}{lll}
 \tableline\tableline
Parameter        & Value & Reference\tablenotemark{b} \\
\tableline
names            & NGC~1569, UGC~3056, Arp~210, VII Zw 16 &  \\
right ascension  & $04\h 30\m 49\s$ (J2000)               & 1 \\
declination      & $64\degr 50\arcmin 53\arcsec$ (J2000)  & 1 \\
galactic coordinates & l=143\fdg68, b=+11\fdg24       & 1 \\           
adopted distance & 2.0~Mpc                                & 2, 3 \\
inclination      & $63\degr$                              & 4\\
optical velocity & $-73$~\kms\ (heliocentric)             & 5 \\
metallicity       & 0.23 $Z_{\odot}$                       & 6 \\
star formation rate\tablenotemark{a} & $0.5 \ldots 3$ \Msun\ yr$^{-1}$ & 7 \\
                 & $4 \ldots 20$ \Msun\ yr$^{-1}$\ kpc$^{-2}$ & 7 \\
galaxy group     & IC~342/Maffei                          & 8 \\
close neighbors  & UGCA~92 (proj.\ distance: 1\fdg2)      & 8 \\
\tableline
total \HI extent & 9\farcm0\ $\times$ 5\farcm5 ($5.2$~kpc $\times$ $3.2$~kpc) & Fig.~10 \\
                  & (at $N_{\rm HI}=1 \cdot 10^{20}$~cm$^{-2}$ level) & \\
\HI position angle & 112\degr\ (major axis)               & Sect. 3\\
\HI radial velocity  & $\approx -80$~\kms\ (heliocentric) & Sect. 2, see also 4 \\
\HI mass          & $7 \times 10^7\,M_{\odot}$          & Sect. 2 \\
 \tableline
\end{tabular}
\tablenotetext{a}{in the observed field of $\sim$0.14~kpc$^2$; value refers to 
the latest starburst phase and depends on the slope of the initial mass 
function}
\tablenotetext{b}{References: (1) Cotton, Condon, \& Arbizzani (1999), 
(2) \citet{israel88}, (3) \citet{makarova03}, (4) \citet{stil02}
(5) \citet{falco00}, (6) \citet{gonzalez97}, 
(7) \citet{greggio98} (8) \citet{karach94} } 
  \label{basics}
\end{center}
\end{table}

\begin{table}
\begin{center}
 \caption{Main features of the \HI disk of NGC~1569}
\begin{tabular}{lccclc}
 \tableline\tableline
Name & $\alpha_{2000}$ & $\delta_{2000}$ & $\Delta v$ & Peculiarity & Figure \\
    & [$\h\ \m\ \s\ $] & [$\degr\ \arcmin\ \arcsec\ $] & [\kms ] & & \\
 (1) & (2)             & (3)             & (4)        & (5)        & (6) \\
 \tableline
  \HI disk         & 04 30 50 & 64 50 50 & $-152 \ldots >-21$\tablenotemark{a} 
& & \ref{mom0}\tablenotemark{b}\\
  chimney           & 04 30 49 & 64 50 59 & $-75 \ldots -39$\tablenotemark{c} 
& $N_{\rm HI} \approx 4 \cdot 10^{21}$~cm$^{-2}$ &  \ref{mom0}\tablenotemark{d}
 \\
   disturbed center & 04 30 51 & 64 50 55 & $-100 \ldots -50$ 
& $v \approx$ const.\ at $r<450$~pc &  \ref{pv1} \\
  \HI arm          & 04 30 25 & 64 51 25 & $-118 \ldots -64$ 
& west of disk, different kine- &  \ref{mom0}\tablenotemark{e} \\
& & & & matical component & \\
  southern blob    & 04 30 49 & 64 49 09 & $-90 \ldots -57$ 
& south of disk &  \ref{mom0} \\
  junction         & 04 30 56 & 64 49 23 & $-85 \ldots -67$\tablenotemark{c}  
& connects s. blob and disk &  \ref{mom0} \\
  protrusion       & 04 30 46 & 64 49 22 & $-67 \ldots -31$\tablenotemark{c}  
& velocity comparable to spot  &  \ref{mom1}\\
& & & & at southern edge of the disk & \\
  hot spot         & 04 30 40 & 64 51 12 & $-131 \ldots -72$ 
& $\sigma_{\rm v} > 25$~\kms, &  \ref{mom2}\\
& & & & velocity crowding & \\
 \tableline
\end{tabular}
\tablenotetext{a}{channel maps at high radial velocities suffer strongly from 
 Galactic contamination}
\tablenotetext{b}{3~rms cutoff in high-resolution cube}
\tablenotetext{c}{exact velocity range uncertain}
\tablenotetext{d}{coincident with SSCs A and B} 
\tablenotetext{e}{west of optical disk, \Ha and X-ray halo}
\tablecomments{The main features in the \HI distribution and kinematics of the 
 high-resolution cube. The columns are 
 (1) name of the feature, 
 (2) right ascension (J2000), 
 (3) declination (J2000), 
 (4) velocity range, in which the feature is clearly detectable ($> 3$~rms) 
     in the high-resolution channel maps, 
 (5) peculiarity of the feature, (6) figure that shows the feature most clearly
\label{features}}
\end{center}
\end{table}

\begin{table}
\begin{center}
 \caption{List of the halo features}
\begin{tabular}{lccccccc}
 \tableline\tableline
 Name & $\alpha_{2000}$ & $\delta_{2000}$ & $v_0$ & $\Delta v$ & A & S & 
 $M_{\rm HI}$ \\
 & [$\h\ \m\ \s\ $] & [$\degr\ \arcmin\ \arcsec\ $] & [\kms ] & [\kms ] & [kpc $\times$  kpc] & [Jy\,\kms] & [$10^6$ M$_{\odot}$]  \\
 (1) & (2) & (3) & (4) & (5) & (6) & (7) & (8) \\
 \tableline
C   & 04 32 09 & 64 49 25 & $-138$                  & $-90 \ldots -193$ &  2.5 $\times$ 2.4  & 5.64 &  5.3 \\
B1  & 04 31 34 & 64 47 25 & $-116$                  & $-77 \ldots -144$ &  1.5 $\times$ 1.4  & 1.41 &  1.3 \\
B2  & 04 31 17 & 64 47 18 & $-100$                  & $-64 \ldots -134$ &  1.1 $\times$ 1.1  & 1.35 &  1.3 \\
B3  & 04 31 01 & 64 47 05 & $ -78$\tablenotemark{a} & $-51 \ldots -121$ &  1.0 $\times$ 1.3  & 1.10\tablenotemark{a} &  1.0\tablenotemark{a} \\
S1  & 04 30 47 & 64 45 40 & $ -70$\tablenotemark{a} & $-51 \ldots -82$  &  1.3 $\times$ 0.9  & 0.23\tablenotemark{a,b} &  0.2\tablenotemark{a,b} \\
S2  & 04 30 24 & 64 46 14 & $ -71$\tablenotemark{a} & $-51 \ldots -82$  &  1.4 $\times$ 1.0  & 0.27\tablenotemark{a} &  0.2\tablenotemark{a} \\
S3  & 04 30 11 & 64 47 39 & $ -73$\tablenotemark{a} & $-51 \ldots -82$  &  1.4 $\times$ 0.9  & 0.10\tablenotemark{a} &  0.1\tablenotemark{a} \\
S4  & 04 29 56 & 64 48 52 & $ -79$\tablenotemark{a} & $-51 \ldots -108$ &  1.9 $\times$ 0.9  & 0.69\tablenotemark{a} &  0.6\tablenotemark{a} \\
A   & 04 30 02 & 64 50 45 & $ -79$\tablenotemark{a} & $-51 \ldots -126$ &  2.1 $\times$ 1.2  & 2.49\tablenotemark{a} &  2.3\tablenotemark{a} \\
Bn  & 04 31 49 & 64 52 24 & $-123$ & $-51 \ldots -167$ & (2.0 $\times$ 2.1)\tablenotemark{c} & 1.56 & (1.5)\tablenotemark{c}\\
 \tableline
\end{tabular}
\tablenotetext{a}{spectrum may be truncated due to blanking of channels containing 
 Galactic foreground emission}
\tablenotetext{b}{located at the position of the negative side lobe of NGC~1569,
                  emission may be underestimated}
\tablenotetext{c}{if at a distance of 2.0~Mpc}
\tablecomments{This table summarizes the properties of small groups of \HI 
clouds seen in the very deep \HI map (Fig.~\ref{streamoverlay}).
The columns are 
 (1) name: C = companion, B = bridge, S = stream, A = \HI arm, Bn = northern blob, 
 (2) right ascension (J2000), 
 (3) declination (J2000), 
 (4) velocity of the peak emission, 
 (5) integrated velocity range, 
 (6) area of the integrated region, 
 (7) integrated flux, 
 (8) \HI mass.
  \label{tabstream}}
\end{center}
\end{table}

\end{document}